\documentclass[a4paper,11pt]{article}
\usepackage{pos}

\title{Transverse spherocity dependence of azimuthal anisotropy in heavy-ion collisions \\at the LHC using a multi-phase transport model}

\author*[a]{Neelkamal Mallick}
\author[a,b]{Raghunth Sahoo}
\author[c]{Sushanta Tripathy}
\author[d]{Antonio Ortiz}

\affiliation[a]{Department of Physics, Indian Institute of Technology Indore, Simrol, Indore 453552, India}
\affiliation[b]{CERN, CH 1211, Geneva 23, Switzerland}
\affiliation[c]{INFN - sezione di Bologna, via Irnerio 46, 40126 Bologna BO, Italy}
\affiliation[d]{Instituto de Ciencias Nucleares, Universidad Nacional Aut\'onoma de M\'exico, M\'exico Distrito Federal 04510, M\'exico}

\emailAdd{Neelkamal.Mallick@cern.ch}
\emailAdd{Raghunath.Sahoo@cern.ch}
\emailAdd{Sushanta.Tripathy@cern.ch}
\emailAdd{Antonio.Ortiz.Velasquez@cern.ch}

\abstract{One of the event shape observables, the transverse spherocity ($S_0$), has been studied successfully in small collision systems such as proton-proton collisions at the LHC as a tool to separate jetty and isotropic events. It has a unique capability to distinguish events based on their geometrical shapes. In our work, we report the first implementation of transverse spherocity in heavy-ion collisions using a multi-phase transport model (AMPT). We have performed an extensive study of azimuthal anisotropy of charged particles produced in heavy-ion collisions as a function of transverse spherocity ($S_0$). We have followed the two-particle correlation (2PC) method to estimate the elliptic flow ($v_2$) in different centrality classes in Pb-Pb collisions at $\sqrt{s_{\rm NN}}= 5.02$~TeV for high-$S_0$, $S_0$-integrated and low-$S_0$ events. We found that transverse spherocity successfully differentiates heavy-ion collisions’ event topology based on their geometrical shapes, i.e., high and low values of spherocity. The high-$S_0$ events have nearly zero elliptic flow, while the low-$S_0$ events contribute significantly to the elliptic flow of spherocity-integrated events.
}

\FullConference{%
 
 The Ninth Annual Conference on Large Hadron Collider Physics - LHCP2021\\
 7-12 June 2021\\
 Online
}

\begin{document}
\maketitle

\section{Introduction}
The production of a hot and dense, deconfined state of matter known as the quark-gluon plasma (QGP) has already been established in heavy-ion collisions at the Large Hadron Collider (LHC) at CERN, Switzerland, and Relativistic Heavy Ion Collider (RHIC) at BNL, USA. Recent studies at the LHC show heavy-ion-like features such as ridge-like structures~\cite{CMS:2016fnw} and strangeness enhancements~\cite{ALICE:2016fzo} in \textit{pp} collisions. To understand the system dynamics and production of jets in \textit{pp} collisions, an event shape observable, known as the transverse spherocity ($S_0$), has been introduced recently at the LHC~\cite{Cuautle:2014yda, Ortiz:2015ttf, Salam:2010nqg, Bencedi:2018ctm,ALICE:2019dfi}. These studies show that transverse spherocity ($S_0$) has unique capabilities to distinguish events based on its geometrical shapes \textit{i.e.} jetty and isotropic events. The study of transverse spherocity in heavy-ion collisions may reveal new and unique physics results where the formation of QGP is already known. This study in heavy-ion collisions shall also complement the current event shape approach based on flow vector analysis at the LHC~\cite{ATLAS:2015qwl,Masera:2009zz}.

In this work~\cite{Mallick:2020ium}, we report the first implementation of transverse spherocity in heavy-ion collisions using a multi-phase transport (AMPT) model~\cite{Lin:2004en}. We have performed an extensive study of azimuthal anisotropy of charged particles produced in Pb-Pb collisions at $\sqrt{s_{\rm NN}} = 5.02$~TeV as a function of transverse spherocity ($S_0$). We have followed the two-particle correlation (2PC) method to estimate the elliptic flow ($v_2$) and subtract non-flow from our calculations by following standard experimental procedures. 

\begin{figure}[h]
\begin{center}
\includegraphics[scale=0.6]{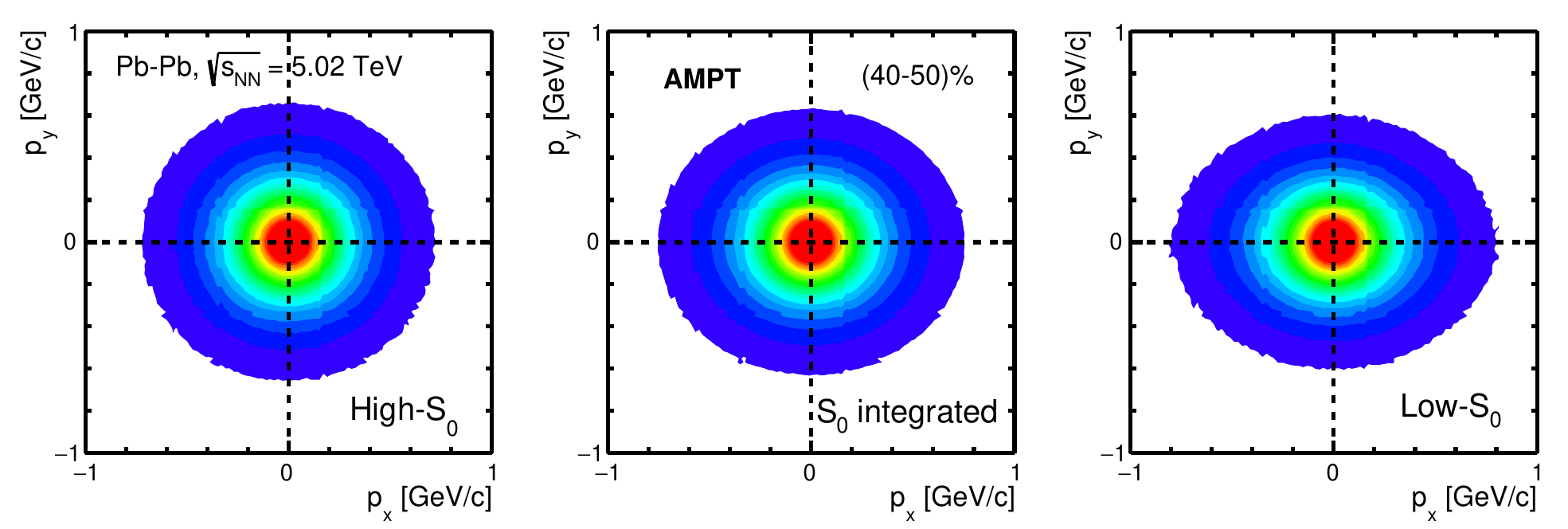}
\caption{(Color Online) Transverse momentum space correlation ($p_{\rm y}$ vs. $p_{\rm x}$) for different spherocity classes in (40-50)\% central Pb-Pb collisions at $\sqrt{s_{\rm NN}}=5.02$ TeV using AMPT model. (Fig. 2~\cite{Mallick:2020ium})}
\label{pypx}
\end{center}
\end{figure}

Transverse spherocity ($S_0$) is an event property which is also a collinear and infrared safe quantity~\cite{Cuautle:2014yda,Ortiz:2015ttf,Salam:2010nqg} and defined as follows,
\begin{eqnarray}
S_{0} = \frac{\pi^{2}}{4} \bigg(\frac{\Sigma_{i}~|\vec p_{T_{i}}\times\hat{n}|}{\Sigma_{i}~p_{T_{i}}}\bigg)^{2}.
\label{eq1}
\end{eqnarray}
Here,  $\hat{n} (n_{T},0)$ is a unit vector known as the spherocity axis, which minimizes the ratio in Eq.~\ref{eq1}. For jetty events $S_{0}\approx0$ and for isotropic events $S_{0}\approx1$~\cite{ALICE:2019dfi}. We have used the mid-rapidity ($|\eta|<0.8$) spherocity distribution with $p_{\rm T}>0.15$ GeV/c with events having at least five such tracks to meet similar conditions as in the ALICE experiment at the LHC. The low-$S_0$ and high-$S_0$ events represent the events lying in the bottom 20\% and top 20\% in the $S_0$ distributions whereas $S_0$-integrated events take all events into account.

Figure~\ref{pypx} represents the transverse momentum space correlation ($p_{\rm y}$ vs. $p_{\rm x}$) for high-$S_0$, $S_0$-integrated and low-$S_0$ events in (40-50)\% central Pb-Pb collisions at $\sqrt{s_{\rm NN}}=5.02$~TeV using AMPT model. The elliptic flow in $S_0$-integrated events can be clearly seen from the elliptic shape of the correlation plot. This is indeed credited to the initial pressure gradient caused due to the initial almond-shaped nuclear overlap region in semi-central collisions, which is then translated to the momentum space ($p_{\rm x}>p_{\rm y}$) correlations. The interesting thing is that the high-$S_0$ events show almost spherical momentum correlation indicating the presence of nearly zero elliptic flow in such events. Whereas the events with low-$S_0$ show a greater elliptical shape correlation. That indicates low-$S_0$ events should be more elliptic and therefore contribute more towards $v_2$.  

\section{Results and Discussions}

 To estimate the elliptic flow we have used the two-particle correlation method (2PC)~\cite{ATLAS:2015qwl,Voloshin:2008dg}. The two particle correlation function $C(\Delta\eta, \Delta\phi)$ is constructed by taking the ratios of same-event pairs $S(\Delta\eta, \Delta\phi)$ to mixed-event pairs $B(\Delta\eta, \Delta\phi)$ given by,
 
 \begin{eqnarray}
C(\Delta\eta ,\Delta\phi) = \frac{S(\Delta\eta ,\Delta\phi)}{B(\Delta\eta ,\Delta\phi)}.
\label{eq2}
\end{eqnarray}

\begin{figure}[h]
\includegraphics[scale=0.24]{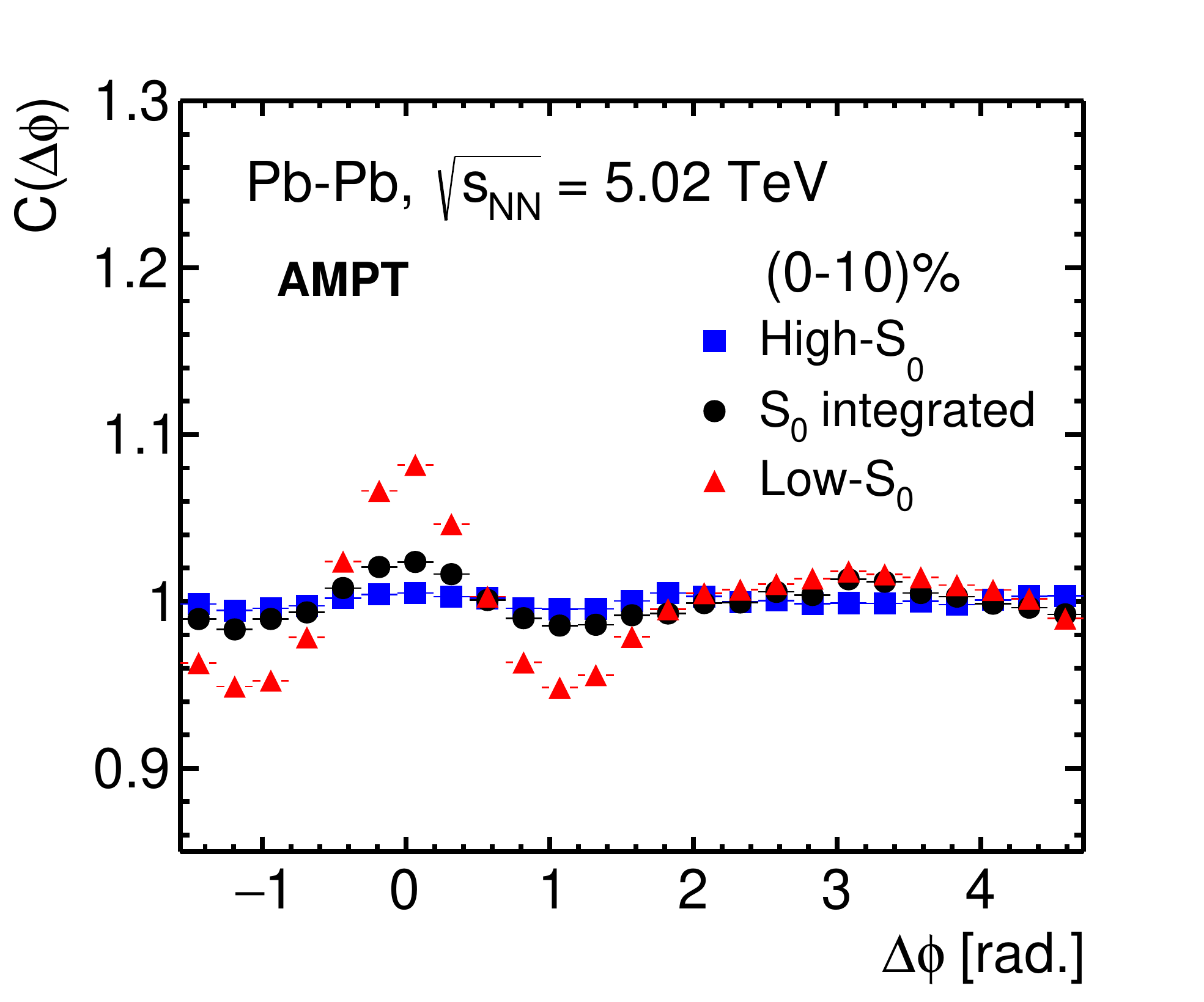}
\includegraphics[scale=0.25]{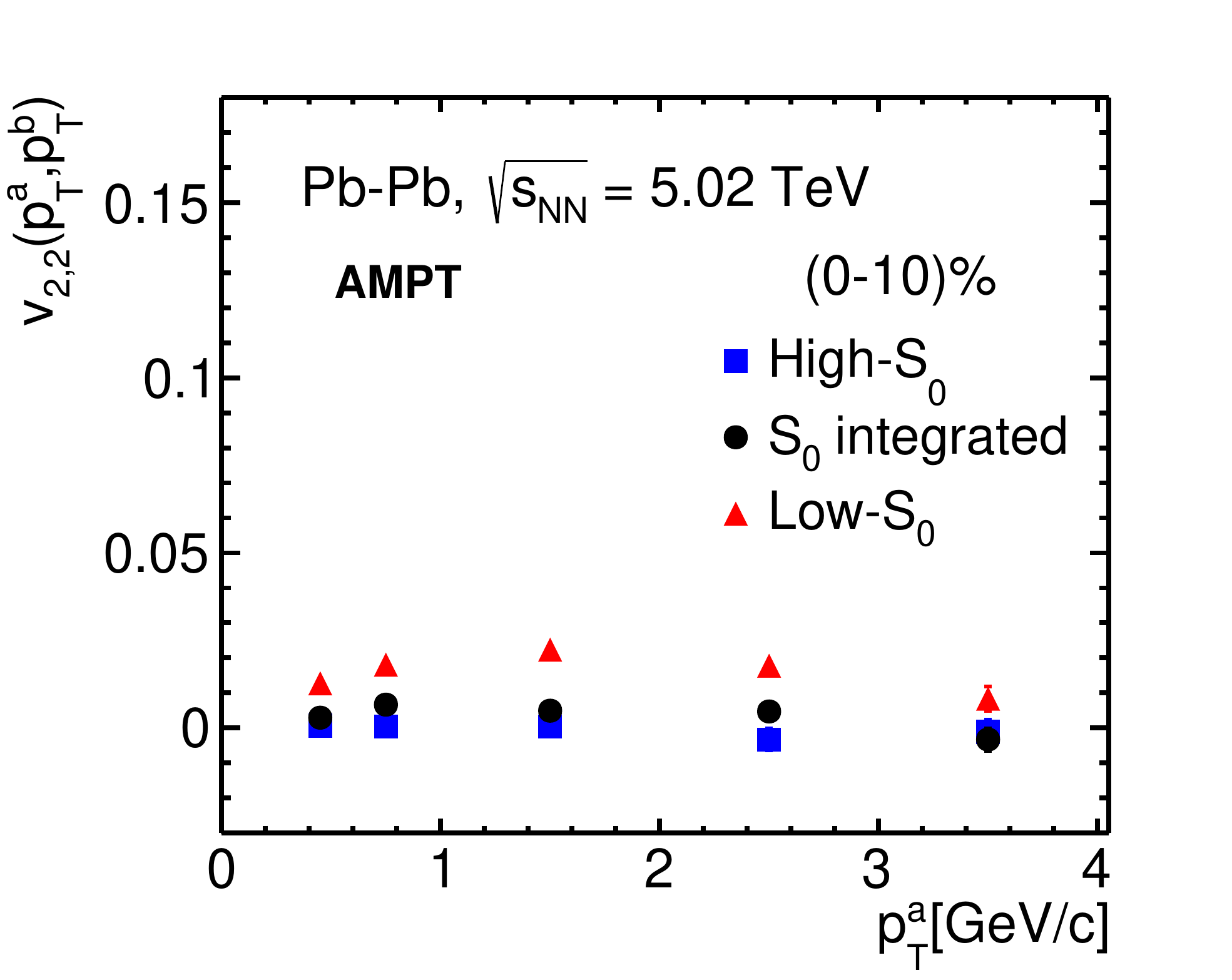}
\includegraphics[scale=0.25]{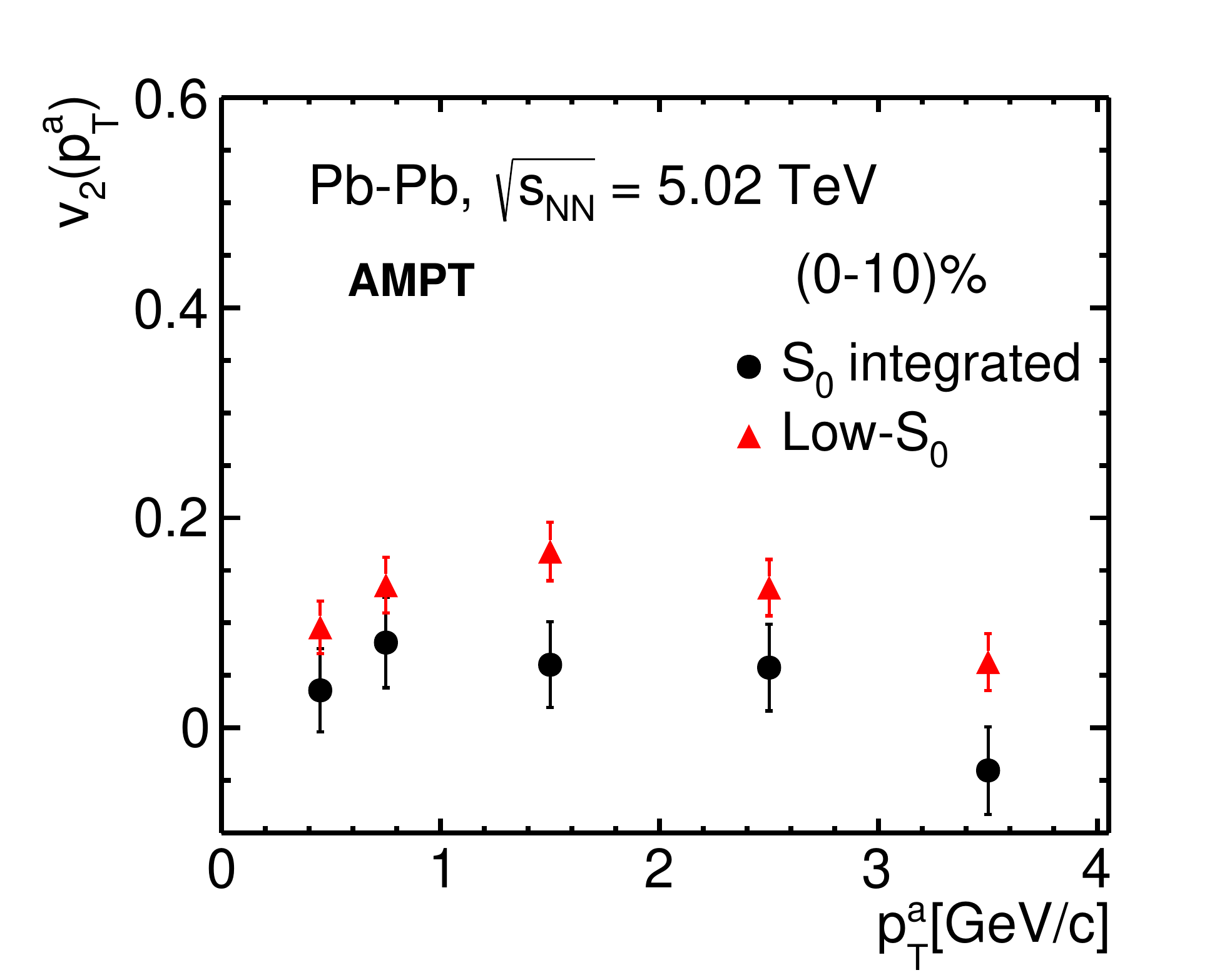}
\includegraphics[scale=0.24]{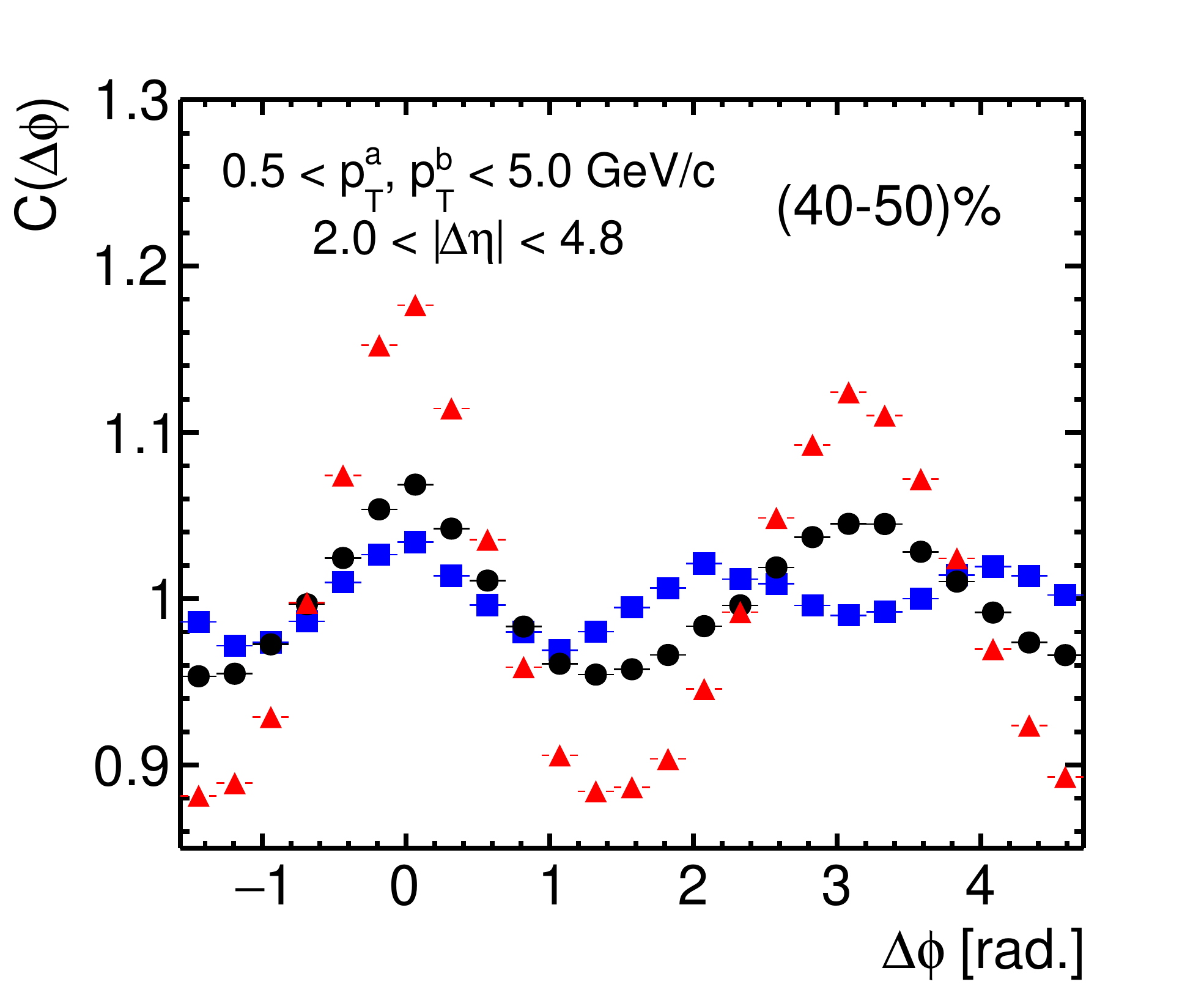}
\includegraphics[scale=0.25]{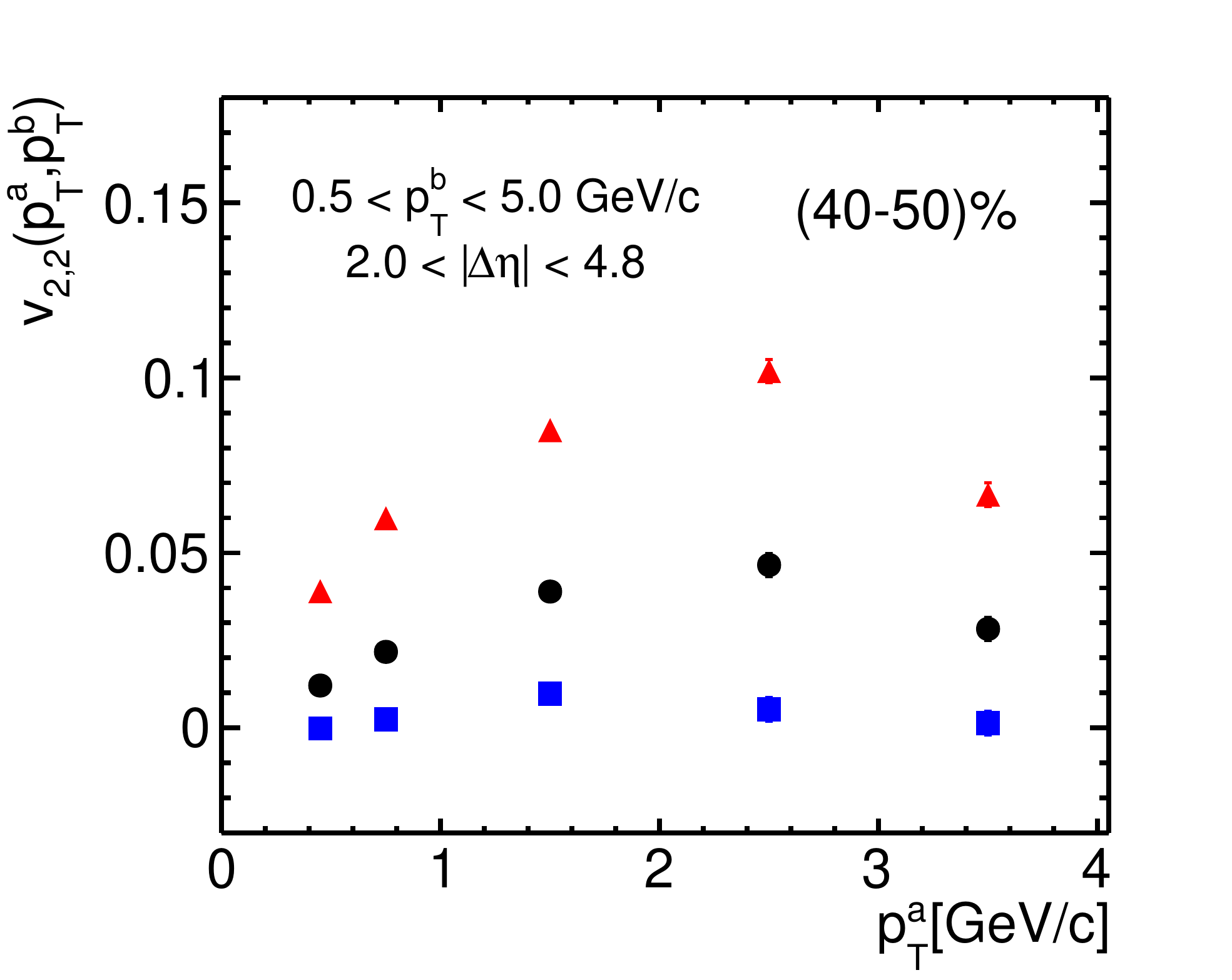}
\includegraphics[scale=0.25]{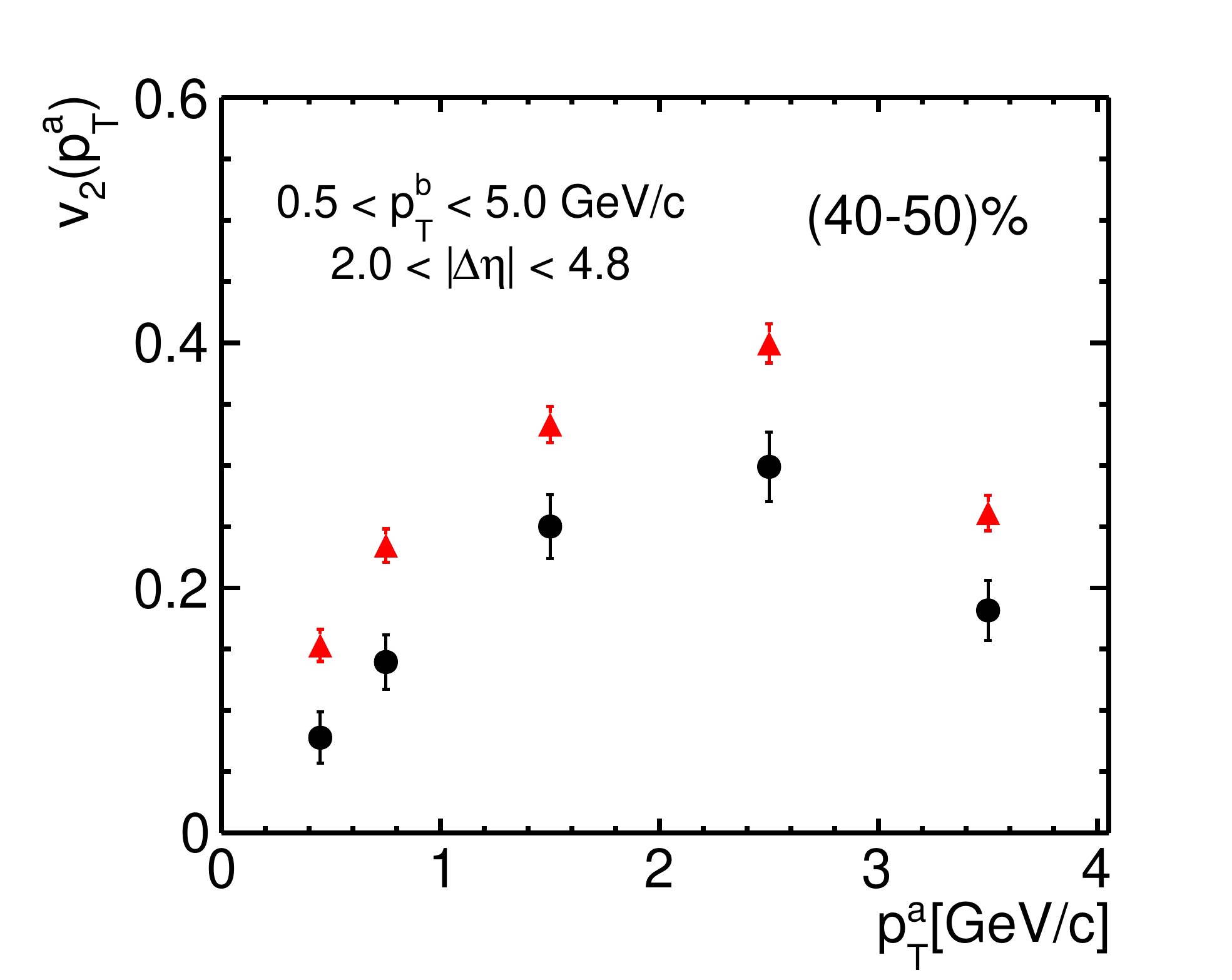}
\caption{(Color Online) First column: One dimensional azimuthal correlation of charged particles, second column: two particle elliptic flow coefficient ($v_{2,2}(p_{T}^{a},p_{T}^{b})$) of charged particles as a function of $p_{T}^{a}$, and  third column:  single particle elliptic flow coefficient ($v_{2}(p_{T}^{a})$) of charged particles as a function of $p_{T}^{a}$ for low-$S_0$, high-$S_0$ and $S_0$-integrated events in Pb–Pb collisions at $\sqrt{s_{\rm NN}}= 5.02$ TeV for 0\%–10\% (top), 40\%–50\% (bottom) centrality classes using AMPT model. (Figs. 4-6~\cite{Mallick:2020ium})}
\label{fig2}
\end{figure}

The one dimensional correlation function ($C(\Delta\phi)$) is then calculated by integrating all the pairs with pseudorapidity gap $2.0<|\Delta\eta|<4.8$. This step helps in reducing the contributions from non-flow effects~\cite{ATLAS:2015qwl}. From the $C(\Delta\phi)$ distribution, the two particle flow coefficient ($v_{2,2}(p_{\rm T}^a,p_{\rm T}^b)$) could be easily obtained as~\cite{ATLAS:2012at,CMS:2013wjq,ALICE:2011svq}, 

\begin{eqnarray}
v_{n,n} (p_{\rm T}^{a},p_{\rm T}^{b}) = \langle cos(n\Delta\phi) \rangle = \frac{\sum_{m=1}^{N} cos(n\Delta \phi_m) \times C(\Delta \phi_m)}{\sum_{m=1}^{N} C(\Delta \phi_m)}.
\label{3}
\end{eqnarray}
where, $N$ = 200 is the number of $\Delta\phi$ bins in the range $-\pi/2 < \Delta \phi < 3\pi/2$. Here, $v_{n,n}$ are symmetric functions with respect to $p_{\rm T}^{a}$ and $p_{\rm T}^{b}$. The single particle flow coefficient could be obtained as,

\begin{eqnarray}
v_{n,n}(p_{\rm T}^{a},p_{\rm T}^{b})= v_{n}(p_{\rm T}^{a}) v_{n}(p_{\rm T}^{b}).
\label{eq8}
\end{eqnarray}
Figure~\ref{fig2} first column shows that the shape and nature of the 1D azimuthal correlation vary with spherocity, and one can observe that the amplitude of the correlation is greater in low-$S_{0}$, intermediate in $S_0$-integrated and lower in high-$S_0$ events. Again, the amplitude of the correlation is more in semi-central collisions compared to most central collisions. 
Figure~\ref{fig2} second column represents the two particle elliptic flow co-efficient $v_{2,2}(p_{T}^{a},p_{T}^{b})$ as a function of spherocity. Here, $v_{2,2}(p_{T}^{a},p_{T}^{b})$ has strong centrality dependence with showing greater values towards semi-central collisions. As far as spherocity is concerned, the high-$S_0$ events are found to have nearly zero elliptic flow while the low-$S_0$ events contribute significantly to elliptic flow of spherocity-integrated events. This is evident from the third column of Fig.~\ref{fig2}.

\section{Summary}
In summary, we found that transverse spherocity successfully differentiates heavy-ion collisions’ event topology based on their geometrical shapes, {\em i.e.} high and low values of spherocity ($S_0$). The high-$S_0$ events are found to have nearly zero elliptic flow, while the low-$S_0$ events contribute significantly to the elliptic flow of spherocity-integrated events. Transverse spherocity is anti-correlated to elliptic flow.
\section*{Acknowledgements}
R.S. acknowledges the financial supports under the CERN Scientific Associateship and the financial grants from DAE-BRNS Project No. 58/14/29/2019-BRNS. The authors would like to acknowledge the usage of resources of the LHC grid computing facility at VECC, Kolkata, and the computing farm at ICN-UNAM. S.T. acknowledges the support from INFN postdoctoral fellowship in experimental physics. A.O. acknowledges the financial support from CONACyT under Grant No. A1-S-22917.

\end{document}